\documentclass[iop]{emulateapj}
\usepackage[flushleft]{threeparttable}
\usepackage{booktabs}
\usepackage{array}

\slugcomment{Accepted to ApJL}

\shorttitle{}
\shortauthors{Batalha et al.}

\begin{document}

\title{Strategies for Constraining the Atmospheres of Temperate \\ Terrestrial Planets 
with JWST}

\author{Natasha E. Batalha, Nikole K. Lewis, Jeff Valenti, Kevin Stevenson} 

\affil{Space Telescope Science Institute, Baltimore, MD, 21218}

\author{Michael R. Line}
\affil{School of Earth \& Space Exploration, Arizona State University, Phoenix, AZ 85282}

\begin{abstract}
\emph{TESS} is expected to discover dozens of temperate terrestrial planets orbiting M dwarfs whose atmospheres could be followed up with the James Webb Space Telescope (\emph{JWST}). Currently, the TRAPPIST-1 system serves as a benchmark to determine the feasibility and resources required to yield atmospheric constraints. We assess these questions and leverage an information content analysis to determine observing strategies for yielding high precision spectroscopy in transmission and emission. Our goal is to guide observing strategies of temperate terrestrial planets in preparation for the early \emph{JWST} cycles. First, we explore \emph{JWST}’s current capabilities and expected spectral precision for targets near the saturation limits of specific modes. In doing so, we highlight the enhanced capabilities of high efficiency readout patterns that are being considered for implementation in Cycle 2. We propose a partial saturation strategy to increase the achievable precision of \emph{JWST}'s NIRSpec Prism. We show that \emph{JWST} has the potential to detect the dominant absorbing gas in the atmospheres of temperate terrestrial planets by the 10th transit using transmission spectroscopy techniques in the near-IR. We also show that stacking $\gtrapprox$10 transmission spectroscopy observations is unlikely to yield significant improvements in determining atmospheric composition. For emission spectroscopy, we show that the MIRI LRS is unlikely to provide robust constraints on the atmospheric composition of temperate terrestrial planets. Higher precision emission spectroscopy at wavelengths longward of those accessible to MIRI LRS, as proposed in the \emph{Origins Space Telescope} concept, could help improve the constraints on molecular abundances of temperate terrestrial planets orbiting M-dwarfs. 
\end{abstract}


\section{Introduction}
There are currently four Earth-sized (0.9~R$_\oplus<$R$_p<$1.5~R$_\oplus$) planets in the habitable zone of their host stars that are amenable to atmospheric follow-up with the James Webb Space Telescope (\emph{JWST}): TRAPPIST-1d,e,f and LHS 1140b \citep{gillon2016temperate, gillon2017seven,dittmann2017temperate, grimm2018nature}. The Transiting Exoplanet Survey Satellite (\emph{TESS}), slated for launch in 2018, is expected to find dozens more of these temperate Earth-sized planets orbiting cool, nearby stars \citep{sullivan2015transiting}. In order to prepare for this new era of exoplanet characterization, studies have sought to assess the feasibility of detecting the atmospheres of temperate worlds, and to determine optimal observing strategies for yielding these constraints.

One strategy for assessing the feasibility of characterizing temperate Earth-sized planets is to define a detection criterion and compute the number of transits needed to meet that criterion. For example, \citet{Batalha2015transiting} determine how many transmission spectra must be coadded to detect H$_2$O at a S/N=15. \citet{louie2017simulated} did a comprehensive analysis of the expected S/N return for the full \emph{TESS} planet yield observed via transmission spectroscopy using \emph{JWST}'s NIRISS SOSS. Most recently, \citet{morley2017observing} determine the number of transits needed to detect molecular features in Earth, Venus and Titan-like atmospheres at S/N=5. The consensus of this work indicates that 10$+$ transits must be co-added in order to yield significant S/N on molecular features of small Earth-like planets orbiting M-dwarf stars. In some cases though, up to 100 transits were needed to detect key atmospheric features \citep{morley2017observing}.

Another strategy for assessing the observability of temperate planets is to use sophisticated retrieval algorithms to determine with what fidelity atmospheric properties can be constrained \citep[e.g.][]{benneke2012atmospheric, dewit2013constraining,Barstow2015transit,Greene2016characterizing}. \citet{benneke2012atmospheric} and \citet{Barstow2015transit} simulate observations of a GJ 1214-like system utilizing the NIRSpec prism \citep{dorner2016model}. \citet{dewit2013constraining} simulate observations of an Earth-like planet orbiting an M7V star at 15~pc utilizing the NIRSpec gratings. \citet{Greene2016characterizing} simulate observations of a mini-Neptune with NIRISS Single Object Slitless Spectroscopy (SOSS), the NIRCam long wave grism, and MIRI Low Resolution Spectroscopy (LRS). All of these studies offer insights into the kind of constraints we expect in \emph{JWST}-era spectra of exoplanets, however they do not explore a wide range of planet-types or observing strategies due to the computationally-intensive nature of retrieval algorithms. 

To combat this, information content analysis has been leveraged to suggest ways to optimize the science yield of \emph{JWST} observations of a large variety of planets ranging from warm-Neptunes to hot-Jupiters \citep{batalha2016information,Howe2017information}. \citet{batalha2016information} suggest that the best modes for constraining gas giants exoplanets' terminator temperature, metallicity, and C/O, are the combination of NIRISS SOSS and NIRSpec G395H. This analysis focuses on targets brighter than J=10.5, and therefore does not include the NIRSpec prism (saturates at J=10.5). Here, we extend this analysis toward temperate planets and include an in-depth analysis of the utility of the NIRSpec Prism to explore these worlds. We choose TRAPPIST-1 as a case study, with the goal of generally guiding observing strategies of temperate Earth-sized planets with \emph{JWST}. 

In \S\ref{sec:model} we describe our methodology for modeling instrument systematics, and transmission \& emission spectra. We also describe how we quantify the information content contained within an observation. In \S\ref{sec:results} we provide our results and discussions, and offer concluding remarks in \S\ref{sec:conc}. 
 
\section{Methods} \label{sec:model}
\subsection{Instrument Simulations}
When using \emph{JWST} to probe exoplanet atmospheres with high-precision time-series observations, the precision expected for each observing mode depends on the stellar energy distribution (SED) of the parent star. We first compute the potential systematic noise sources from each instrument mode using \texttt{PandExo}, described in \citet{batalha2017pandexo}. \texttt{PandExo} relies on the \emph{JWST} Exposure Time Calculator engine \texttt{Pandeia} \citep{Pontoppidan2016pandeia} to compute throughputs, realistic point spread functions and other instrumental effects. Both \texttt{PandExo} and \texttt{Pandeia} generally agree with the instrument team's individual noise simulators to better than 10\% \citep{batalha2017pandexo}. 

First, we compute simulations for each time-series spectroscopy mode, using the standard readout patterns offered in Cycle~1. Then, we explore the expected performance for readout patterns and observing strategies that have have been proposed. We bin all of our calculations to a resolving power of R=100 to facilitate direct comparisons between different instruments (later discussed in \S3.1). Lastly, we use a T=2550 K, M/H=0.4, and logg=4.0 Phoenix stellar model \citep{husser2013new} for all calculations in this work because we are using TRAPPIST-1 as a case study. 

\subsection{Transmission \& Emission Spectra}
Extensive theoretical work has been done to assess the climate, habitability, composition and detectability of the planets that transit TRAPPIST-1  \citep{Barstow2016habitable,dong2017atmospheric,bolmont2017water,unterborn2017constraining,wolf2017assessing,turbet2017modelling,morley2017observing}. Nevertheless, a model capable of predicting, \emph{a priori}, the atmospheric composition of these planets from the few known parameters (mass, radius, orbital properties) does not exist. Therefore, many of the predictions of the atmospheric composition of these planets are grounded in Solar System science. For example, \citet{morley2017observing} create an extensive grid of both primary and secondary transit spectra using the elemental ratios of Earth, Titan and Venus with different incident flux levels, surface pressures and albedos for each planet in their study. 

Due to the complexity and quantity of unknown parameters, here we do not aim to produce chemically consistent spectra in composition or temperature-pressure. Instead, in order to obtain estimates for constraints we might expect from a variety of atmospheres, we explore nine simple chemical prescriptions for each planet. This allows us to assess the impact of the quality and spectral coverage of \emph{JWST} data on how well atmospheric parameters can be constrained. Our transmission and emission model is described in \citep{batalha2016information,line2013near,Greene2016characterizing,line2016influence}. 

All nine chemical scenarios considered here are composed of a combination of H$_2$O, CO$_2$, N$_2$ and CH$_4$, based on the dominant molecules in the atmospheres of rocky Solar System bodies and also the dominant molecules in the \citet{morley2017observing} grid. The main difference between the nine scenarios is the background gas: either H$_2$O-rich, N$_2$-rich or CO$_2$-rich. After the background gas, the three remaining species are added in equal quantities at trace levels (1\%, 0.01\% and 0.0001\%). Adding the gases in equal quantities allows any inability to detect a spectral feature to be attributed to data precision, spectral coverage, or masking by the dominant gas. All compositions are uniform with altitude. 

For each planet we explore temperature-pressure profiles, which we assume can be fully described by a 1D profile. For our 1D profiles we use a 5 parameter double-gray analytic formula \citep{guillot2010on, line2013near}, which for weakly irradiated systems approximate to $T_z^4 \approx 0.75*T^4(p+2.0/3.0)$, where $p$, and $T_z$ are the height-dependent pressure and temperature. Using this scaling, we explore surface temperatures consistent with the full range of potential values given an Earth-like composition, and a range of pressures and Bond albedos from \citet{morley2017observing}. This range is particularly important in the analysis of emission spectra. We use a surface temperature of 200 K and 400 K to set the our pessimistic and optimistic atmospheric constraints in emission, respectively. 

Clouds mute or mask the atmospheric features in transmission spectra \citep[e.g.][]{Kreidberg2014clouds, sing2016continuum}. For terrestrial planets, the models of the cloud-microphysics for Earth \citep[e.g.][]{albrecht1989aerosols, tinsley2000influence}, Venus \citep[e.g.][]{knollenberg1980microphysics,allen1984cloud}, and Titan \citep[e.g.][]{mckay2001physical, rannou2006latitudinal} have all been guided by observations. For exoplanets, a general cloud model does not yet exist. Therefore, we use a grey opacity source at two different pressures levels (0.01 and 0.1 bars) to set our optimistic and pessimistic cases. Optimistically, we assume observations would be limited by the tropopause of the planet, located at 0.1 bars in the Solar System planets \citep{robinson2014common}. Pessimistically, we assume observations would be limited by the formation of high altitude clouds in slowly rotating habitable zone planets \citep{kopparapu2017habitable}. For emission, we do not include the presence of clouds because their reduced optical depths have less of an impact on the spectrum \citep{fortney2005effect}.
\subsection{Information Content Theory} 
\citet{batalha2016information} details our information content methodology. We describe the relevant sections here. The information content is a quantity that describes how the state of knowledge of a system has increased (relative to the prior) by making a measurement \citep{shannon2001mathematical, line2012information}. Here, we are specifically interested in the posterior covariance matrices, $\mathbf{\hat{S}}$, which describe the uncertainties on each of the state vector parameters. We assume that our atmospheric state is described by $T$, $\xi_{H_{2}O}$, $\xi_{CO_{2}}$, $\xi_{CH_{4}}$, $\xi_{N_{2}}$, and $\times R_p$. $T$ is the temperature above the tropopause which we set to planet's $T_{eq}$, $\xi_i$ is the concentration of the $i$th gas, and $\times R_p$ is a factor to account for the radius arbitrarily set at 10 bars. $\mathbf{\hat{S}}$, can be computed as
\begin{equation}
    \mathbf{\hat{S}} = (\mathbf{K^TS_e^{-1}K + S_a^{-1}})^{-1}
\end{equation}
where $\mathbf{K}$ is the Jacobian matrix, which describes the model sensitivity, $\mathbf{S_a}$ is the \emph{a priori} covariance matrix, and $\mathbf{S_e}$ is the error covariance matrix. We assume that the observer has no prior information so that $\mathbf{K^TS_e^{-1}K >> S_a^{-1}}$. This ensures that our calculations are driven by the model sensitivity and the \emph{JWST} data, not the prior. 

\begin{figure*}[ht]
\centering
 \includegraphics[angle=0,width=0.7\linewidth]{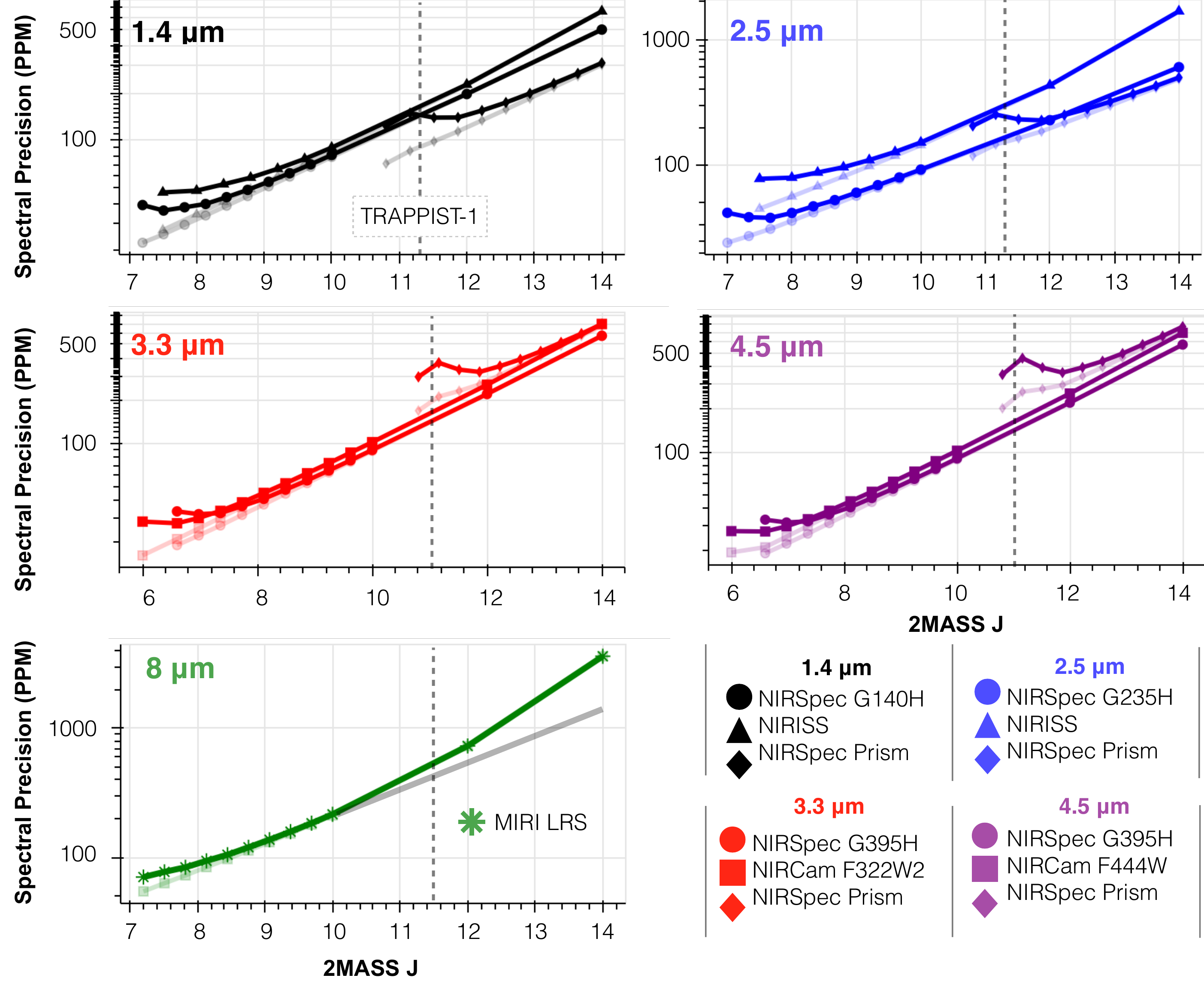}
\caption{Curves show the spectral precision on the planet spectrum as a function of J magnitude at various wavelengths. Each simulation is composed of 2 hours of total observing time. Colored opaque lines represent the \emph{Reset} and \emph{read} mode and transparent lines are for a 100\% efficient observation. The dashed vertical lines represent the J-magnitude of TRAPPIST-1, for reference. In the $8\mu$m panel, the grey line represents the pure-photon limited precision. All simulations are binned to R=100 for comparison. \textbf{Main Points}: 1) Low observing efficiency limits precision for bright targets. 2) NIRSpec Prism becomes dominated by read noise at longer wavelengths. 3) MIRI is background limited past J=10\label{fig:noise}}
\end{figure*}

\section{Results \& Discussion}\label{sec:results}
\subsection{Instrument Systematics \& Observing Strategies} \label{sec:noise}
\emph{JWST} nondestructively reads charge in a pixel as it accumulates during an integration. Each integration begins with a \emph{reset} frame, during which pixels in the subarray are reset one at a time. For time series science, each downlinked \emph{group} consists of one \emph{frame}, the result of reading each pixel once. The first \emph{frame} after a \emph{reset} will usually be used to establish a zero point, or bias level. Subsequent \emph{frames} will be read accumulated science photons. The efficiency of an observation is then =$\frac{n-1}{n+1}$, where $n$ is the number of groups. If the bias level is known a priori, then efficiency is $\frac{n-1}{n}$. The MIRI detectors are somewhat more efficient at small $n$ because it can read then reset pixels in a single frame time.

For bright targets near the saturation point of the instrument, this readout pattern becomes inefficient. The opaque lines in Figure~\ref{fig:noise} show our results for the expected spectral precision for each mode as a function of J magnitude at key wavelengths (H$_2$O at 1.4 $\mu$m \& 2.5 $\mu$m, CH$_4$ at 3.3 $\mu$m, CO/CO$_2$ at $\sim$4.5$\mu$m).

Each opaque curve follows the expected photon-limited relationship until the lower limit in magnitude, where the detectors begin to approach saturation. This flattening out in precision is the result of decreasing the groups within an integration for brighter targets, which decreases the observing efficiency to 33\% for the brightest targets. The transparent lines in Figure \ref{fig:noise} show the expected spectral precision if \emph{JWST} had 100\% observing efficiency at all magnitudes within the saturation limits of the instrument. For temperate terrestrial planets, which require very high spectral precision, we need to determine strategies to increase this observing efficiency for targets near the saturation limits of a given mode.
\begin{figure*}[ht]
\centering
 \includegraphics[angle=0,width=0.7\linewidth]{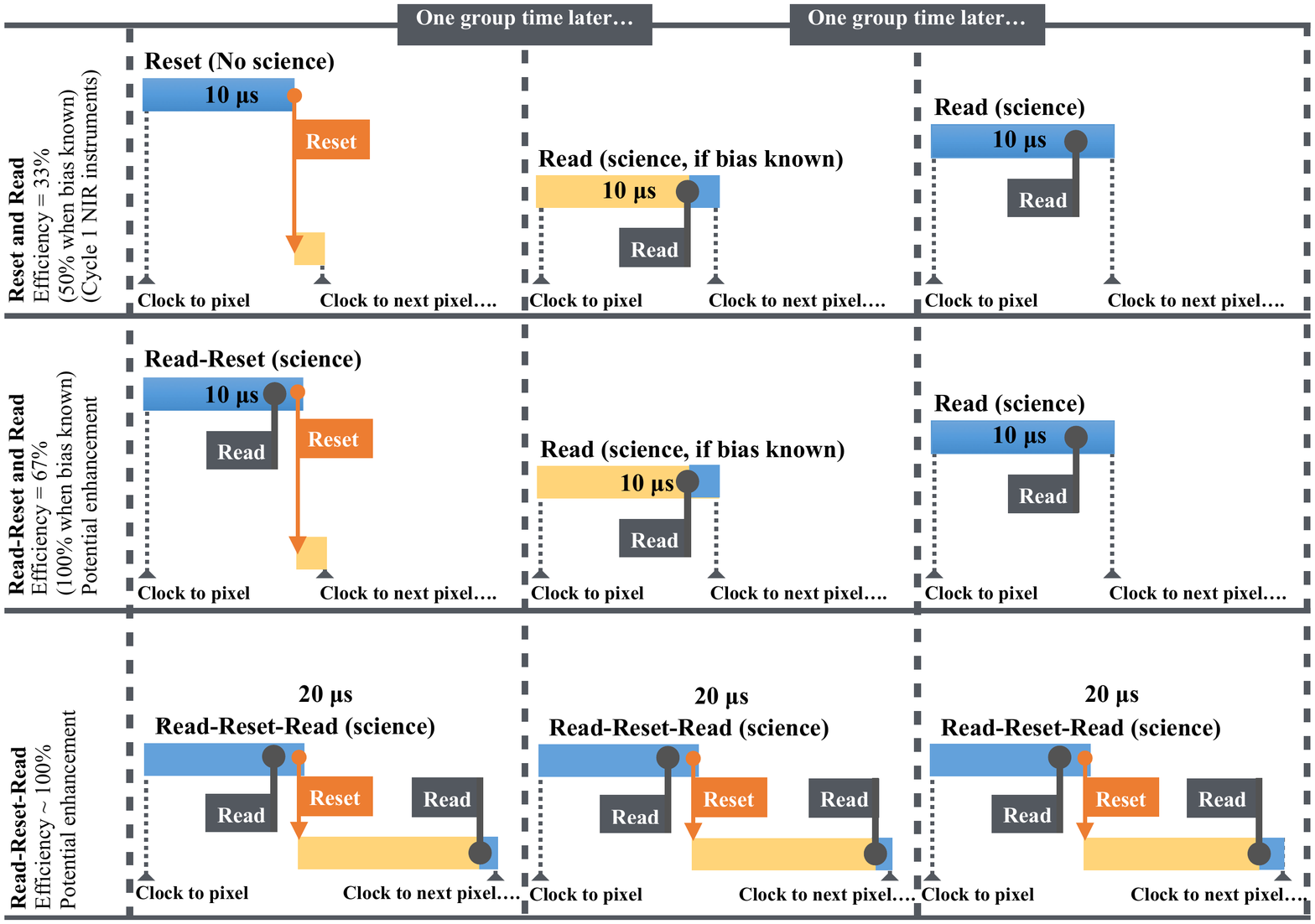}
\caption{The top panel shows the presently supported readout pattern for Cycle~1. The middle and bottom panels show potential enhancements. Each of the three columns (separated by a dashed vertical line) represent one \emph{group} time. Blue always represents science time, yellow represents potential bias time. The main difference between the top and middle panel is the \emph{read} that occurs immediately before the \emph{reset} in the first column. See \S3.1 for a more thorough explanation. \textbf{Main Point:} In Cycle~1, readout patterns will limit the observing efficiency for targets near the saturation point of a particular instrument to 33\%. Enhanced readout patterns could change this to $\sim$100\% efficiency. \label{fig:rmode}}
\end{figure*}
Therefore, high efficiency readout patterns are under investigation for NIRISS and NIRSpec, shown in Figure \ref{fig:rmode}. Each of the three columns in Figure \ref{fig:rmode} shows what takes place during a single pixel's read and/or reset. As stated above, a full \emph{group} is the result of clocking through all the pixels one time. The top panel is the currently supported readout pattern, and the bottom two show potential enhancements.  In the \emph{Read-Reset and Read} pattern, the position of the reset frame is moved to take place directly after an individual pixel is read (middle panel). Reading each pixel immediately \emph{before} the reset measures all accumulated charge and yields 100\% efficiency, only if the reset level is known. If the reset level is not known, the \emph{Read-Reset-Read} readout pattern would have to be implemented to yield $\sim$100\% efficiency (bottom panel). 

If these new readout patterns get implemented, it will greatly increase efficiency of targets near the saturation limits of specific high-precision time-series modes. However, these enhanced readout patterns are not currently slated to be available for observations in JWST’s Cycle~1. TRAPPIST-1, J=11.3, saturates the NIRSpec Prism after the third group. Therefore, an observation with no saturation leads to $\frac{3-1}{3+1}\sim0.50$ observing efficiency across the entire detector. Using the NIRSpec Prism is advantageous because it yields a 1-5 $\mu$m spectrum in a single transit. For targets not accessible with the Prism, observers will have to decide if they want to split up observations between modes to yield an entire 1-5 $\mu$m spectrum, or use all their time in a single observing mode. 

One additional caveat with the NIRSpec Prism, shown in Figure \ref{fig:noise}, is its decrease in precision toward longer wavelengths. At wavelengths less than 2.5 $\mu$m, the NIRSpec Prism attains higher spectral precision than other available instruments at identical 2MASS J. The NIRSpec Prism becomes less favorable at higher wavelengths though, because the stellar SED dramatically drops off, causing read noise to dominate over photon noise, and because the efficiency of the observation is limited by saturation at the shorter wavelengths. 

To combat both these sources of decreased precision, we propose an observing strategy to partially saturate the detector. \emph{JWST} acquires sampled up the ramp data and will return a data product for every single group within the integration. Therefore, unless the observation is saturated at the end of the second group, a variable number of groups can be used to extract the full wavelength space, regardless of saturation.

\begin{figure}
\centering
 \includegraphics[angle=0,width=3in]{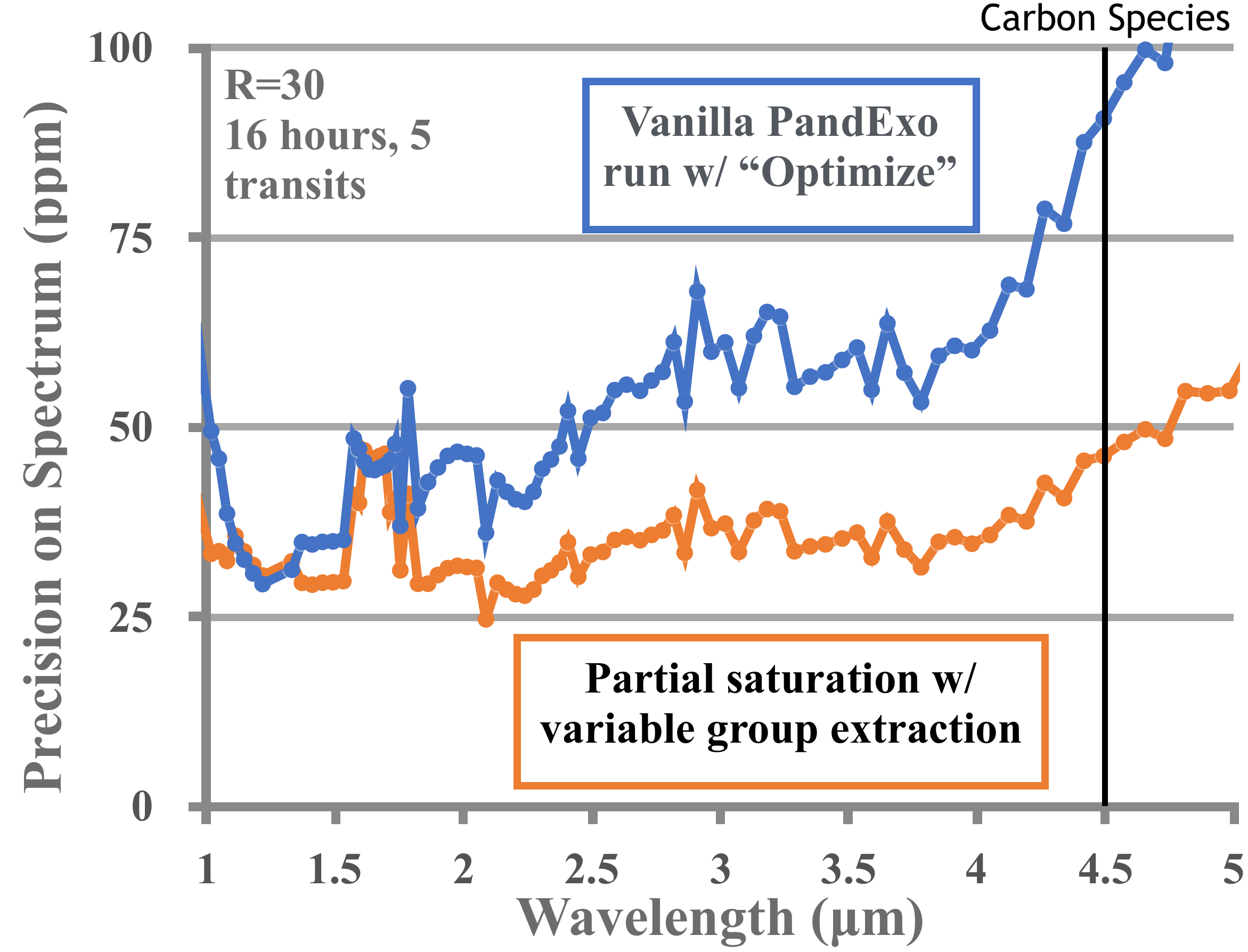}
\caption{A comparison of two different observing strategies with the NIRSpec Prism. The blue curve shows the result of a \texttt{PandExo} run with the number of groups determined by the ``optimize'' option (ngroup=3). The orange curve shows the result of ngroup=6 for the total exposure time. \textbf{Main Point:} Partial saturation can increase precision for NIRSpec Prism observations of bright targets. \label{fig:prism}}
\end{figure}

Figure \ref{fig:prism} shows the spectral precision of this variable group observing strategy (orange), versus a non-saturated \texttt{PandExo} run with 3 groups (blue) for the TRAPPIST-1 system. In this particular phase space (targets with low-efficiency observations), this proposed strategy has the potential to increase the precision at longer wavelengths by a factor of 2 for the NIRSpec Prism. Note, this observing strategy has not yet been formally introduced into the \texttt{PandExo} package.

In order to obtain emission spectroscopy of cool planets, observations in the mid-IR, using MIRI LRS, will be required. There are currently no plans to increase the observing efficiency of MIRI LRS near the saturation limit. There are other ways to increase observing efficiency that are not specific to the exoplanet case. Although, because the MIRI detectors are read out more efficiently than the near-IR detectors, this is less problematic (see Figure \ref{fig:noise}). One caveat of MIRI is that observations become background limited past J=10, seen by the grey photon-limited line in the MIRI LRS panel of Figure \ref{fig:noise}. Therefore, TRAPPIST-1 is not an ideal target to study emission spectroscopy of terrestrial exoplanets and it will be especially important for \emph{TESS} to detect planets with J$<10$ to optimize observations with MIRI LRS. 

\subsection{Transmission Analysis}
The relationship between the uncertainties on the retrieved parameters and the number of transmisison spectra needed are similar for each planet in the TRAPPIST-1 system. The differences between planets are driven by differences in temperature and gravity, which set the strength of the molecular features through the scale height $=kT/\mu g$. We choose TRAPPIST-1f to illustrate our results. TRAPPIST-1f is the outermost habitable zone planet (219 K), and has a gravity similar to that of Earth (8.33 m/s$^2$). 
\begin{figure*}[ht]
\centering
 \includegraphics[angle=0,width=5in]{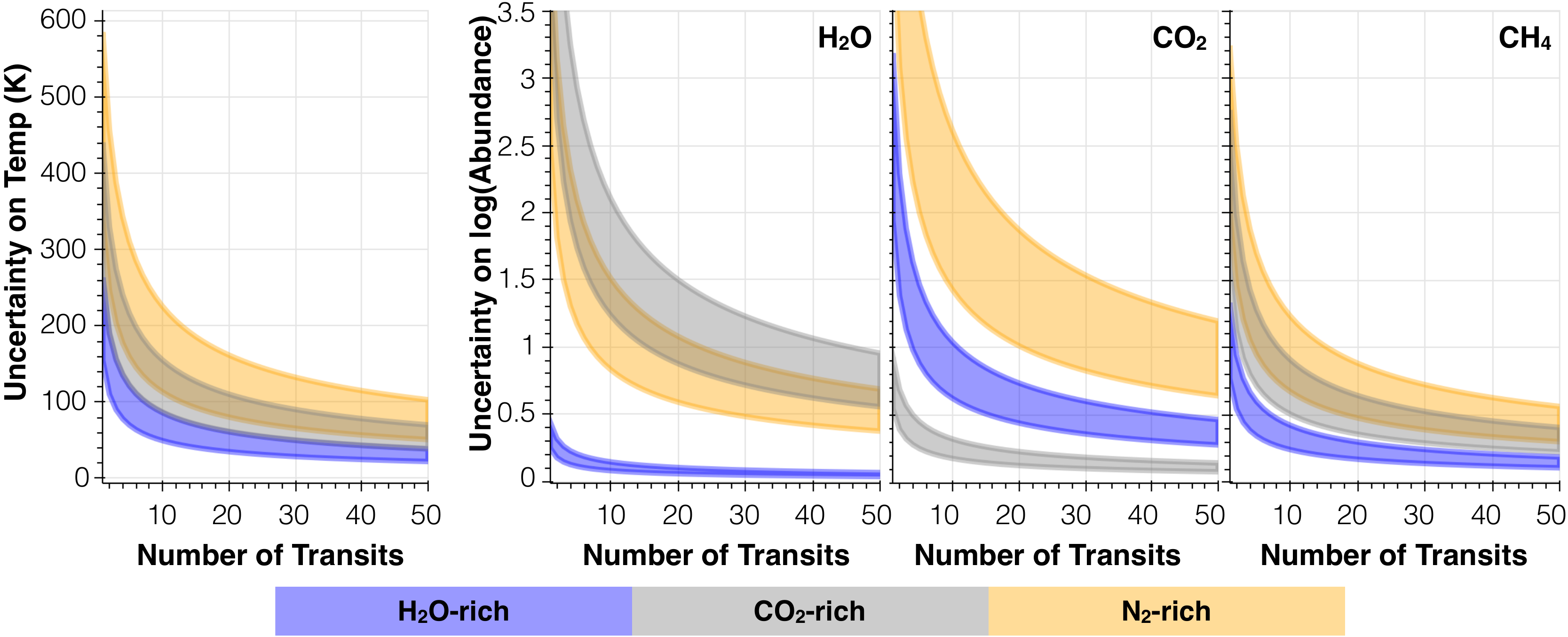}
\caption{Uncertainties on each state vector parameter calculated from the information content analysis for TRAPPIST-1f. Each transit observation consists of a NIRSpec Prism observation with total observation time = 4 $\times t_{14}$. Purple curves correspond to a H$_2$O-rich atmosphere with 0.01\% of CO$_2$, CH$_4$, and N$_2$. Orange curves correspond to an N$_2$-rich atmosphere with 0.01\% of CO$_2$, CH$_4$, and H$_2$O. And grey curves correspond to a CO$_2$-rich atmosphere with 0.01\% of N$_2$, CH$_4$, and H$_2$O. The upper and lower bound of the curve are set by the optimistic (P=0.1 bar) and pessimistic (P=0.01 bar) specifications for the grey cloud top pressure.\textbf{Main Points:} 1) The greatest gain  in information occurs in the first 10 transits, 2) The dominant molecular absorber is detected after the 10th transit in all cases, 3) Transit transmission spectroscopy will not constrain temperature profiles \label{fig4}}
\end{figure*}

Figure \ref{fig4} shows the expected uncertainty on the atmospheric parameters of TRAPPIST-1f after each transit, if it were observed with NIRSpec Prism's partial saturation strategy in transmission. The upper and lower bound of the curve is set by the pessimistic (P=0.01 bar) and optimistic (P=0.1 bar) specifications for the grey cloud top pressure, respectively. When there is no information content in the observation, the uncertainty approaches the prior (12 dex for abundances and 1000 K for temperatures). We do not show the results for detecting N$_2$ because it is void of molecular features unless temperatures are very high \citep{schwieterman2016identifying}. Therefore, it cannot be directly detected or constrained. 

Temperature is difficult to constrain in transmission spectroscopy, regardless of atmospheric composition. Meaningful constraints ($<\pm 50$~K) are only achievable with 10$+$ transits. For abundances, in all our cases, the dominant \emph{absorber} is constrained by the 10th transit. Therefore for the TRAPPIST-1 system, if no atmospheric signals are detected by the 10th transit, it is unlikely that co-adding more would unveil new information. However, if the dominant absorber is detected by the 10th transit, additional observations could reveal trace gases in the atmosphere at the 0.01\% level.  

Our results demonstrate the high potential the NIRSpec Prism has for detecting a wide variety of molecular features. H$_2$O has dominant absorption features from 1-2$\mu$m, CH$_4$ has dominant absorption features from 3-4 $\mu$m, and CO$_2$ has dominant absorption features from 4-5 $\mu$m. In Cycle~1, it is important to survey this entire parameter space. For targets too bright to be accessible with NIRSpec Prism, observations with a combined NIRISS SOSS and NIRSpec G395H observation yield higher information content results, despite the lower precision that comes from splitting time between two modes. 
\begin{figure*}[ht]
\centering
 \includegraphics[angle=0,width=5in]{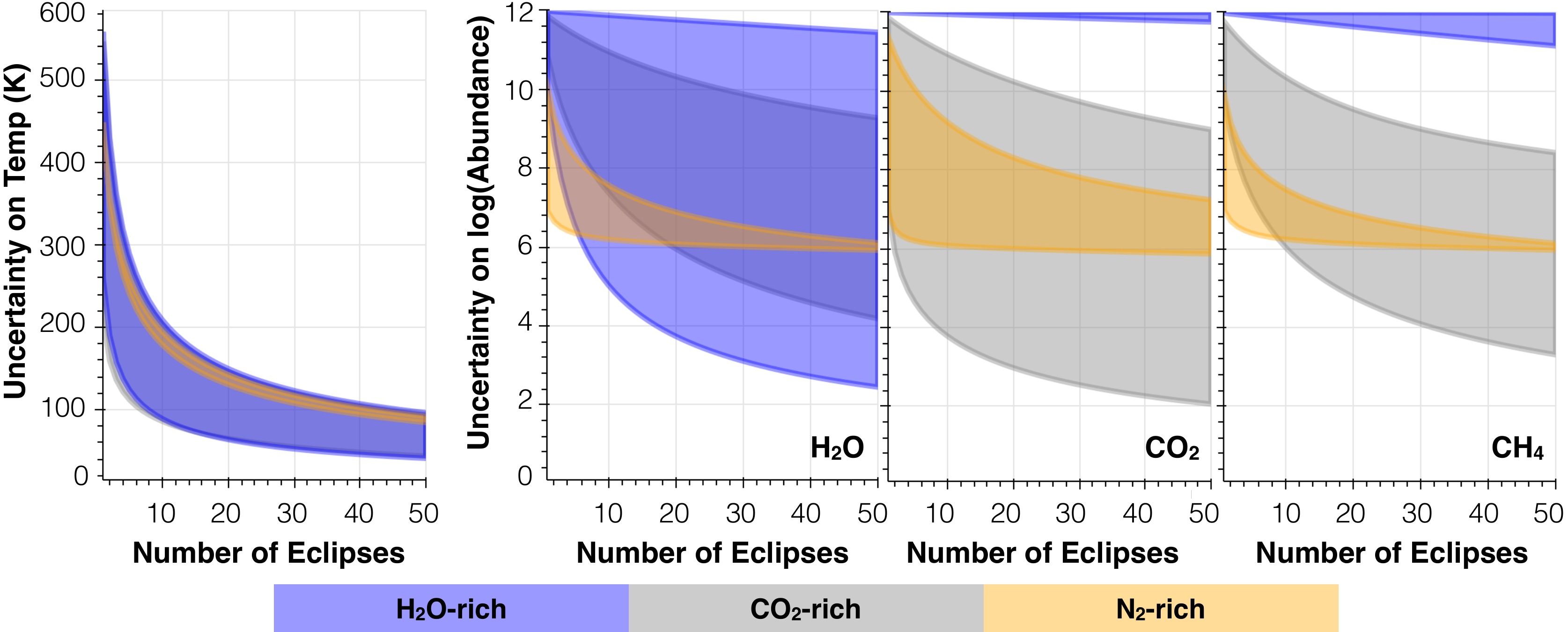}
\caption{Same as Figure \ref{fig4}, but for emission spectroscopy with MIRI LRS. Here, the upper and lower bound of the curve are set by the optimistic (T=400 K bar) and pessimistic (T=200 K) specifications for the surface temperature at 100 and 1 bar, respectively. \textbf{Main Point:} Emission spectra of temperate planets with \emph{JWST}'s MIRI LRS is unlikely to provide strong atmospheric constraints of truly temperate planets. Future facilities, such as the \emph{Origins Space Telescope} concept, could provide the required precision and wavelength coverage needed to improve these constraints. \label{fig5}}
\end{figure*}
\subsection{Emission Analysis} 
For emission we also show the case of TRAPPIST-1f. Figure \ref{fig5} shows the constraints on the atmospheric state vector parameters. Emission spectroscopy is more sensitive to the atmospheric temperature structure than transmission spectroscopy. However, the uncertainties on temperature for emission spectroscopy (Figure \ref{fig5}) are comparable to that of transit transmission (Figure \ref{fig4}). This is because the \emph{JWST} MIRI LRS data does not have sufficient precision to detect the small signal that comes from the emission of temperate planets at 5-12$\mu$m. The constraints on abundances are also highly driven by the prior (12 dex). No molecules are constrained within 1 dex for less than 50 transits. Detecting molecular features in emission spectroscopy of truly temperate exoplanets will be very difficult with \emph{JWST}'s MIRI LRS. This conclusion is also supported by the analysis of \citep{morley2017observing}, which suggests photometry of temperate planets as an alternative to emission spectroscopy. 

\section{Conclusions}  \label{sec:conc}
Here, we used \texttt{PandExo} in combination with an information content analysis to determine optimal strategies for constraining the atmospheres of the planets in the TRAPPIST-1 system-- with the ultimate goal of guiding observations of temperate terrestrial planets. We summarize our conclusions below:
\begin{itemize}
    \item Bright targets near the saturation point of the specific instrument mode have low observing efficiency. This is especially true of observations of the TRAPPIST-1 system with NIRSpec Prism, which is a  favorable mode because of its ability to get a complete spectrum (1-5 $\mu$m) in one transit. The Prism also is dominated by read noise at longer wavelengths because of this low efficiency and because the stellar SED drops towards 5 $\mu$m. While high efficiency read modes are being investigated, we outline a partial saturation strategy for the NIRSpec Prism that can increase observing efficiency and decrease the effect of readnoise at long wavelengths. 
    \item Using a partial saturation strategy with the Prism, we will detect the dominant atmospheric absorber of temperate terrestrial planets by the 10th transit. If we do not detect anything by the 10th transit, it is not likely that coadding more transits would unveil more information. If we do detect the dominant absorber by the 10th transit, more transits could reveal trace gases at the 0.01\% level.
    \item Emission spectroscopy with MIRI LRS is unlikely to provide strong atmospheric constraints of truly temperate (surface temperatures=200-400 K) planets. Future missions/facilities, such as the \emph{Origins Space Telescope} concept, could provide the required wavelength coverage and precision to provide robust constraints on the atmospheres of temperate terrestrial worlds in the mid-IR. 
\end{itemize}

\acknowledgments
We thank the members of STScI's STARGATE team for their comments and discussions. Specifically, we thank Hannah Wakeford, Jonathan Fraine, and Giovanni Bruno for their helpful feedback. We also thank Eddie Bergeron for investigating more efficient detector readout modes for bright targets. NB, NK, and JV acknowledge support from Grant NNX15AC86G from NASA/GSFC for the JWST Telescope Scientist Investigation.

Facilities: \facility{JWST}

\end{document}